\documentclass[twocolumn,superscriptaddress]{revtex4}

\usepackage{amsmath}  
\usepackage{amsfonts} 
\usepackage{graphicx} 

\begin{document}


\title{Estimating the Moon to Earth radius ratio with a smartphone, a telescope and an eclipse}

\author{Hugo Caerols}
\email{hugo.caerols@uai.cl}
\affiliation{Facultad de Ingenier\'{\i}a y Ciencias, Universidad Adolfo Ib\'a\~nez, Santiago 7941169, Chile.}

\author{Felipe A. Asenjo}
\email{felipe.asenjo@uai.cl}
\affiliation{Facultad de Ingenier\'{\i}a y Ciencias, Universidad Adolfo Ib\'a\~nez, Santiago 7941169, Chile.}


\date{\today}

\begin{abstract}
On January 20th, 2019, a total lunar eclipse was possible to be observed in Santiago, Chile. Using a smartphone attached to a telescope, photographs of the phenomenon were taken. With  Earth's shadow on those images, and using textbook geometry, a simple open-source software and analytical procedures, we were allowed to calculate the ratio between the radii of the Moon and the Earth. The results are in very good agreement with the correct value for such ratio. This shows the strength of the smartphone technology to get powerful astronomical results in a very simple way and in a very short amount of time.
\end{abstract}
 
\maketitle 

From ancient times, the different features of planets and moons have created a huge interest \cite{helden}. Aristarchus was one of the first  to study the relative relations among Earth, Moon and Sun \cite{Heath,Momeni}.
This interest has remained until today, and therefore, it is always relevant to make this knowledge more appealing to the younger generations. 
 Nowadays, smartphone technology has become an important tool to teach  physics, and this gives us a huge opportunity to bring science closer  to students in a simpler manner. In this work, we show how simple photographs of a partial lunar eclipse are sufficiently good  to estimate the ratio between the Moon and the Earth radii. After taking the photographs, the  procedure for the calculation is straightforward and it can be reproduced easily in an one--hour class. 

The first total lunar eclipse of the year 2019, ocurred on the January 20th. We took several photographs of this eclipse from Santiago of Chile. Among all the photographs,  the best ones are those in which the border line of the Earth shadow is narrow, which are 17 in total. 
Examples of these photographs are shown in Figs.~\ref{centroluna}, \ref{centrosombra} and \ref{completo}. Also,  the figures  show the procedure to obtain the ratio between Moon and Earth's shadow radii (explained in the following sections).

Although several estimations of the ratio between  Moon and  Earth radii have been proposed before \cite{otherworks,Momeni,Cowley,Oostra, Murphy}, the strength of our work lays in its simplicity and reproducibility by any person with basic knowledge of geometry and algebra. By using these photographs to estimate this ratio, our method offers a very good opportunity to teach simple geometry, focusing in its importance to understand the structure and dynamics of the celestial bodies. As we will show below, it is the combination of our current advanced technology and textbook  geometry what allow us to obtain an excellent estimation for the ratio between the Moon and the Earth  radii.


In the following sections, we detail the procedure for the estimation using the photographs. The 17 taken images allow us to determine the ratio between the Moon and Earth's shadow radi  by constructing in a geometrical manner the radius of each circle. Then, by using the Thales' theorem, we relate the previous ratio with the real physical ratio between Moon and Earth radii. In order to estimate the correct ratio, we develop an analytical optimization scheme to show how accurate our proposal is. Finally, we discuss the limitations of this kind of proposal.

\section{Ratio estimation}

Earth and moon are not perfect spheres. Thus, several radii can be defined. The International Astronomical Union (IAU) \cite{IAU} have defined the Earth's
equatorial radius, i.e. the distance from its center to the equator, to be equal to $6378.1370$ [km]. Similarly, the IAU have determined \cite{Guillermier} the that the mean Lunar Radius $k$ in units of Earth's equatorial radius (i.e. the ratio between the Moon and Earth radii) is 
\begin{equation}\label{valork}
k=0.2725076\, ,
\end{equation}
where the equatorial Moon radius is $1738.09$ [km].

The purpose of this work is to estimate the value \eqref{valork} in the simplest possible way for students with an smartphone and a free-source software. 
A smartphone with a 12 megapixels camera
 was attached by a suitable adapter to a telescope Celestron 114 short, 1000mm focal length, F/9, Model \#31041 with  an ocular  26mm Plossl 1.25''.  Using a moon filter,  photographs of the first total lunar eclipse of 2019 were taken. All photographs are images of a partial lunar eclipse. In the next section we will explain why the following procedure is not good for a full eclipse.

The value \eqref{valork}  of $k$ can be related to the shadow that Earth cast over the Moon in a lunar eclipse in the form
\begin{equation}\label{valork2}
k=\frac{R_M}{R_E}=\left(\frac{R_{sh}}{R_E}\right)\left(\frac{R_M}{R_{sh}}\right)=\mu\, z\, ,
\end{equation}
where $R_M$ is the Moon's radius, and $R_E$ is Earth's radius, and  $R_{sh}$ is the radius of the  Earth's shadow over the Moon.
Here we have defined 
\begin{equation}\label{valork2mu}
\mu=\frac{R_{sh}}{R_E}\, ,\quad z=\frac{R_M}{R_{sh}}\, ,
\end{equation}
such that $\mu$ is the ratio between the Earth's shadow and Earth radii, and $z$ is the ratio between the Moon and Earth's shadow radii. 
The parameter $\mu$ measures the size of Earth's shadow casted over the Moon at a given distance from Earth at the eclipse time. It measure the reduction of Earth's radius at the distance of Moon. 
On the hand, the parameter $z$ measures the effective ratio between that shadow and Moon's radius. The photographs taken with the telescope and the smartphone  allow us to determine the parameter $z$ for each one of the images. Nevertheless, we need to find the value of $\mu$. This parameter has a geometrical nature and it depends on Earth
and Moon positions. 

In the following sections, we calculate each of these parameters, and prove  that we are able to estimate $k$ with a very good accuracy by using them.

\subsection{Determination of $\mu$}

Due to the shadow cone produced by Earth, its shadow has a smaller radius as we move away from it. This implies that we can relate geometrically the sizes between Earth's radius and the radius of its shadow casted over the Moon. This allow us to find $\mu$ from Eq.~\eqref{valork2mu}.
\begin{figure}[h!]
\centering
\includegraphics[width=8.4cm]{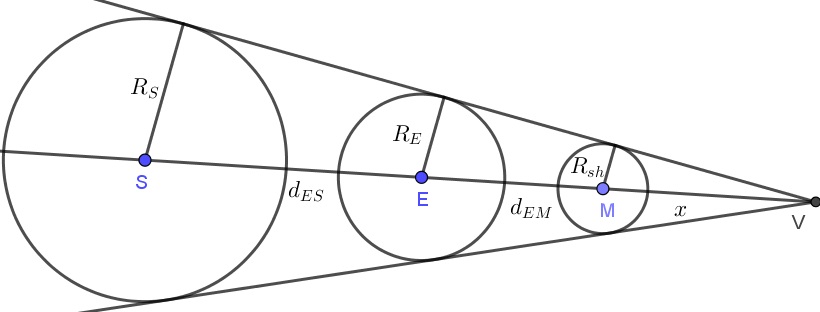}    
\caption{Diagram that shows the relation between  Earth's shadow and Earth radii
 at the Moon distance.}
\label{eclipseFigshadowM}
\end{figure}

The geometrical relations are schematically represented in Fig.~\ref{eclipseFigshadowM}. Here,  $R_S$, $R_E$ and $R_{sh}$ represent the Sun, Earth and Earth's shadow radii respectively. Furthermore, $d_{ES}$ (corresponding to the segment $\overline{SE}$) is the distances between the Earth and Sun, while $d_{EM}$ (the segment $\overline{EM}$) is
the distance between the Earth and the Moon at the moment of the eclipse. Lastly, $x$ (segment $\overline{MV}$) is the unknown distance between Moon and the focus of the shadow cone. We can see that all the previous quantities are related by the Thales's theorem, which allow us to find relations between them. These relations are $x/R_{sh}=(x+d_{EM})/R_E$, and $(x+d_{EM})/R_E=(x+d_{EM}+d_{ES})/R_S$.
Solving for $R_{sh}$, we are able to find
\begin{equation}\label{valuofmu}
\mu=\frac{(d_{EM}+d_{ES})R_E-d_{EM}R_S}{d_{ES}\, R_E}\, .
\end{equation}

In order to calculate $\mu$, we need the different values appearing at the right-hand side of the above equation.
During the lunar eclipse on January 20th, 2019, these  values are known \cite{estela}. These are $R_S=695700$[Km] $R_E=6378.137$[Km],  $d_{ES}=1.47208\times 10^{8}$[Km], and $d_{EM}=354171.698$[Km]. Then,  replacing them, we find that
\begin{equation}\label{valemu}
\mu=0.739978\, .
\end{equation}
This value for $\mu$ implies that the Earth's radius is reduced approximately by a factor $0.74$ at the distance of Moon at this particular eclipse at that date. In general, the numerical value \eqref{valemu} for $\mu$ can change. In lunar eclipses,  the Moon passes directly behind Earth which can only occur during full moon phases. However, as the lunar eclipse occurs at a different time of the year, the distances $d_{ES}$ and $d_{EM}$ can  increase or reduce, producing changes in the shadow.

\subsection{Determination of $z$ for each photograph}
\label{deterz}

The parameter $z$ can be calculated using Eq.~\eqref{valork2mu}, by using the ratio between Moon and the Earth's shadow radii in the photographs. One should expect a value near $z=k/\mu=0.3682645$, which results to divide the values \eqref{valork} for $k$ and
 \eqref{valemu} for $\mu$. Thus, our photographs must have $z$ very near to that value.

Using the GeoGebra software \cite{geogebra} (a geometry free-source software) the radii of Moon and Earth's shadow are estimated in terms of units given by the software (distances are measured in centimeters). The units of the software are not important, as we are interested in the dimensionless ratio between the calculated values.

The procedure to obtain the Moon and Earth's shadow radii can be performed using geometry. In general, three points are enough to determine the radius of the circumscribed circumference \cite{cox}. Consider a circumference with undetermined radius. Let us choose three points  in its perimeter. There are two segments that connect these three points. Calculate the mediatrix 
to each segment and you can notice that they intersect. This
allows us to construct two isosceles triangles, that  share one side. Therefore, the intersection point is the center of the circle, and
the sides represent the radius of the circumference circumscribed to the three initial points.

\begin{figure}[h!]
\centering
\includegraphics[width=8cm]{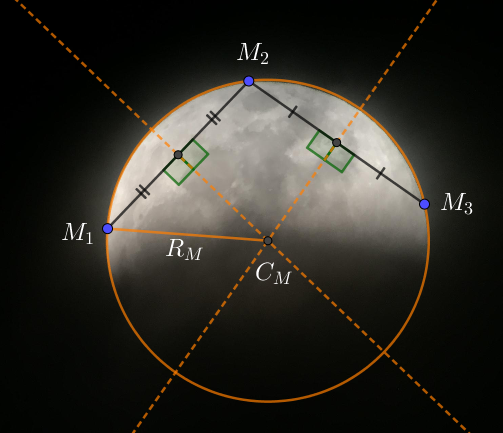}    
\caption{Photography taken from the lunar eclipse. The points $M_i$ ($i=1,2,3$) are on the Moon's surface. These points are used to calculate each mediatrix, and therefore the Moon's radius. This image corresponds to photograph 7 in Table \ref{bosons}.}
\label{centroluna}
\end{figure}
\begin{figure}[h!]
\centering
\includegraphics[width=8cm]{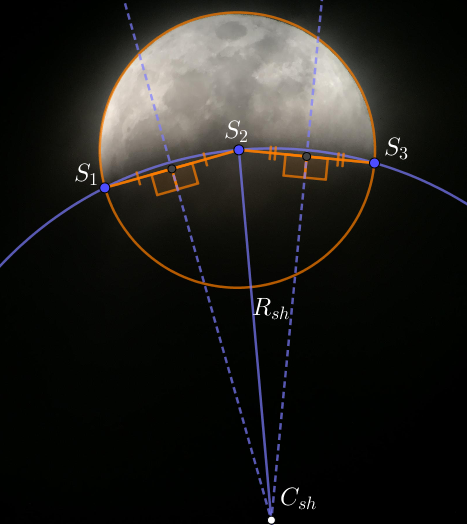}    
\caption{Photography taken from the lunar eclipse.  The points $S_i$ ($i=1,2,3$)  are   on the Earth's shadow perimeter. These points are used to calcute each mediatrix, and therefore the radius of Earth's shadow. This image corresponds to photograph 7 in Table \ref{bosons}. }
\label{centrosombra}
\end{figure}

In Fig.~\ref{centroluna}, the procedure is explained for the Moon. In this case, three points on its surface can be taken in an arbitrary way, called $M_1$, $M_2$ and $M_3$.
Using  GeoGebra software, we can obtain the  mediatrix of segments $\overline{M_1M_2}$  and segment $\overline{M_2 M_3}$. Each mediatrix is depicted in a orange dashed line. The point $C_M$, where the two mediatices intersec, is the center of Moon. Thereby, the segments $\overline{M_1 C_M}$, $\overline{M_2 C_M}$ and $\overline{M_3 C_M}$ are sides of isoceles triangles and equal, being  equivalent to the Moon radius $R_M$. We ask Geogebra to measure that segment, thus obtaining its value in terms of units of the software. 
 We apply the same  approach for the Earth's shadow, depicted in Fig.~\ref{centrosombra}. Using the same photograph than for Fig.~\ref{centroluna}, we choose three points on the Earth's shadow casted on the Moon,  called $S_1$, $S_2$ and $S_3$. Obtaining the mediatrixes of segments $\overline{S_1 S_2}$ and $\overline{S_2 S_3}$ (in blue dashed lines), we are able to find the center of its circumference $C_{sh}$, corresponding to center of Earth's shadow. Anew, the segments $\overline{S_1 C_{sh}}$, $\overline{S_2 C_{sh}}$ and $\overline{S_3 C_{sh}}$ are equal to each other, which define the Earth's shadow radius $R_{sh}$.

\begin{figure}[h!]
\centering
\includegraphics[width=8cm]{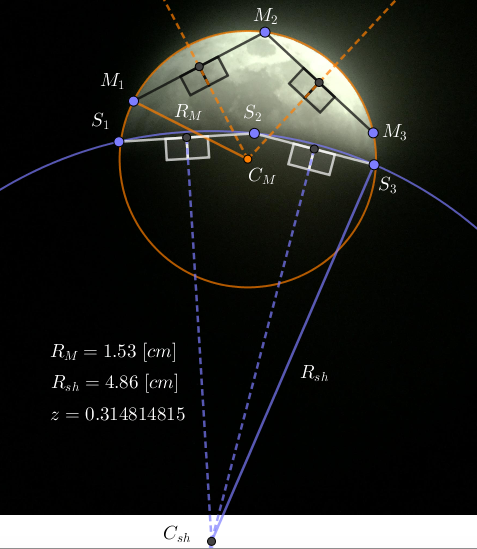}    
\caption{This photography summarizes the complete process to be carried out in any image of the eclipse. This allows to measure $R_M$ and $R_{sh}$, and to determine its ratio $z$ for each image. The image corresponds to photograph 10 of the Table \ref{bosons}.}
\label{completo}
\end{figure}

In Fig.~\ref{completo} the total procedure is shown for photograph 10 of Table \ref{bosons}. Using the above explained procedure, it has been obtained explicitly the radius $R_M=1.53$[cm], and $R_{sh}=4.86$[cm], thus giving  $z=0.314814815$. 

At this point, it is important to remark that we have used this geometrical procedure over the GeoGebra ``Circle through 3 Points" option to fit
any circumference to the chosen three points of Moon and Earth's shadow, as the latter procedure gives the general form for the equation of a circle. In that case, a student needs some background knowledge of analytic geometry in order to determine the radius. We believe that using the geometrical approach, any student will acquire a more intuitive insight of the construction of a circle, requiring at the same time less mathematical background. Thereby, the geometrical approach is available for a wider range of students from different 
levels.

The complete above procedure is applied to 17 photographs of the eclipse.
In Table \ref{bosons} we list the results for all photographs. Each image gives different $R_M$ and $R_{sh}$, and thus different values of $z$. This is due to the variation that can be present as the different photographs are taken.
Every length is obtained in centimeters, software units. However, there is no a problem with this, as we are interested in their dimensionless ratio.
\begin{table}[h!]
\centering
\caption{Calculation of $z$ for 17 photographs}
\begin{ruledtabular}
\begin{tabular}{c c c c }
 Photograph & $R_M$ [cm] & $R_{sh}$ [cm] & $z=R_M/R_{sh}$  \\
\hline	
1 & 1.56 &  5.47 & 0.285191956  \\
2 & 1.56 &  3.96 & 0.393939394 \\
3 & 1.52  & 3.81  & 0.398950131 \\
4 &  1.53 & 3.74  & 0.409090909 \\
5 &  0.84 & 2.62  &  0.320610687\\
6 & 1.51 & 5.4  & 0.27962963 \\
7 & 1.54 & 4.15  & 0.371084337 \\
8 & 0.85 & 3.01  & 0.282392027 \\
9 & 1.54  & 5.78  & 0.266435986  \\
10 & 1.53 & 4.86  & 0.314814815  \\
11 & 0.84 & 2.71  & 0.3099631  \\
12 & 1.55 & 4.86  & 0.318930041 \\
13 & 1.55  &  4.15 & 0.373493976 \\
14 & 1.52 & 3.58  & 0.424581006 \\
15 & 0.86 & 2.3  & 0.373913043 \\
16 & 0.82 & 2.06  & 0.398058252 \\
17 & 0.86 & 2.22  & 0.387387387 
\end{tabular}
\end{ruledtabular}
\label{bosons}
\end{table}

Form the Table \ref{bosons} we can see that the ratio $z$
for each photograph is different. We can calculate the its average value $\bar z$, to get
\begin{equation}\label{barz}
\bar z=\frac{1}{17}\sum_{j=1}^{17} z_j=0.347556863\, ,
\end{equation}
where the index $j$ runs for all the data of Table \ref{bosons}.

The final purpose of the data obtained from the above Table is to use it to obtain the relation 
\eqref{valork2} by taking also into account the value \eqref{valemu} for $\mu$. In the following section, we show how this can be carried out.

\subsection{Determination of predicted ratio $k_e$}

Our data allow us to obtain a predicted ratio $k_e$ between Moon and Earth radii. This predicted ratio must be near the value \eqref{valork}, i.e., $k_e\approx k$, in order to say that this method is very accurate. This is the same that showing that the predicted ratio has the form
\begin{equation}\label{predictedk}
k_e=a\, k\, ,
\end{equation}
for a value of $a$ very near to 1. This is the aim of this section.

If only one photograph of the eclipse is obtained, then with that sole value of $z$, the ratio $k$ can be obtained using Eq.~\eqref{valork2}. However, when several values for $z$ are obtained,  we must develop a scheme to show that they predict a value $a$ close to 1.
In order to do that, let us construct the Error function $E$ defined as
\begin{eqnarray}\label{errorfucntion}
E(a)&=&\sum_{j=1}^{17}\left(a\,  k - \mu z_j \right)^2\nonumber\\
&=&17\, a^2 k^2-2 a k \mu \sum_{j=1}^{17}z_j+\mu^2\sum_{j=1}^{17} z_j^2\, ,
\end{eqnarray}
where $k$ and $\mu$ are the values given in Eqs.~\eqref{valork} and  \eqref{valemu}, respectively. Notice that we are including the definition $k_e=a\,  k$ in that function.
This Error function $E$ depends on $a$, 
and 
it measures how far  the data points $z$ of Table \ref{bosons} are from $k$. If the Error function would result to vanish, then would imply that the relation \eqref{valork2} is exaclty fulfilled, with $a=1$.

\begin{figure}[h!]
\centering
\includegraphics[width=8.4cm]{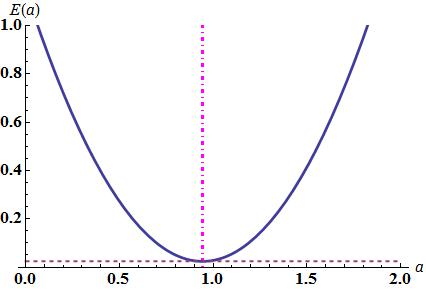}    
\caption{Error function \eqref{errorfucntion}. This plot show that the value \eqref{errorfucntion2} minimize the function $E$.}
\label{errofplot}
\end{figure}

As the Error function \eqref{errorfucntion} depends on $a$, a scheme is required to find its minimum value.
 This allow us to find the value of $a$ that produce the best fit of  relation \eqref{valork2}  to the points $z$ of Table \ref{bosons}. Mathematically, we need to find the minimum value of function \eqref{errorfucntion} that is quadratic for $a$. For a general function in the form $f(a)=\alpha a^2+\beta a+\gamma$ (with $\alpha>0$), its minimum is achieved when $a=-\beta/(2\alpha)$. Using this for the Error function \eqref{errorfucntion}, we find that its minimum occurs in
\begin{equation}\label{errorfucntion2}
a=\frac{\mu}{k}\,  {\bar z}=0.943769378\, ,
\end{equation}
where $\bar z$ is given in \eqref{barz}. This is the same value that can be obtained from more advanced mathetical procedures, where the derivative of $E$ is required to vanish.
This minimum value  can be seen plotting the Error function \eqref{errorfucntion}. In Fig.~\ref{errofplot}, we show the Error function for different values of $a$. The  dashed horizontal line display the minimum value of the Error function evaluated in \eqref{errorfucntion2}, which gives a value of $E=0.0241$. The dot-dashed vertical line shows the value \eqref{errorfucntion2} for $a$. 

Now we are in position to finally calculate the predicted ratio between Moon and Earth radii. Using the value \eqref{errorfucntion2},  the predicted ratio \eqref{predictedk}
has the value
\begin{equation}\label{predictedk2}
k_e=0.257184328\, .
\end{equation}
Compare this value with the IAU value \eqref{valork}. In order to determine how accurate is this estimation, let us calculate the percentage error of our result. This is obtained as
\begin{equation}
\frac{|k-k_e|}{k}=0.05623\, . 
\end{equation}
Consequently, our estimation of $k_e$ contains an error of only 5.6\%, which is a very accurate result.
This shows how powerful is this method to obtain the Lunar radius in units of Earth's equatorial radius, only using photograph taken with an amateur telescope and a smartphone.

\section{Limitations}

It is important to discuss that the above approach has  limitations. Not every photograph is useful, as our method is dependent on the position of the Earth's shadow. This is because the shadow determines the choices for the three points that are used in the construction of the circumferences.
Thereby, there are several issues that need to be considered in order to obtain the best results for the estimations.

\begin{figure}[h!]
\centering
\includegraphics[width=8cm]{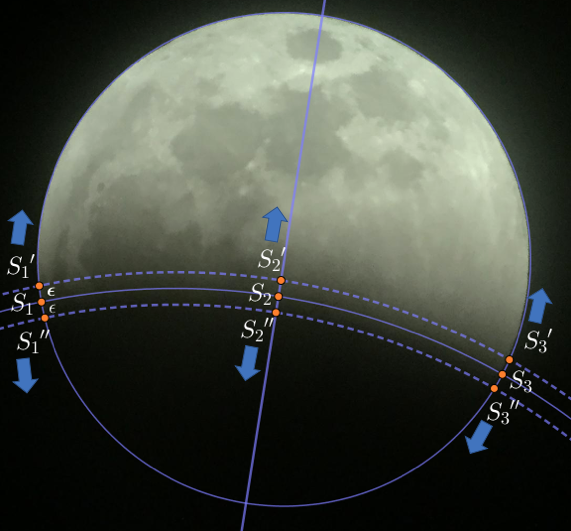}    
\caption{Example of good choices for construction of circunferences using
points delineating the edge of Earth's shadow. The circunferences must be constructed using three  points from the same set ($\{S_1,S_2,S_3\}$, $\{S_1',S_2',S_3'\}$, or $\{S_1'',S_2'',S_3''\}$) in order to mantain an  error of $\epsilon$ order.}
\label{limitation1}
\end{figure}

Any used photograph for this experience must show a well--defined, sharp and narrow edge for the Earth's shadow (a blurry shadow will not work with this method). 
Once the edge is defined,  the position of the points on the perimeter of the Earth's shadow  is a key part of the process. These points should be chosen in the way that they minimize the difference between their radial distances. Otherwise, there is a risk of constructing an ill--defined circle (with a radius approaching to infinity). 
This is explicitly shown in Fig.~\ref{limitation1}, where  an example of three different set of points along the edge of Earth's shadow is used to discuss this issue.
In this example, the set $\{S_1,S_2,S_3\}$ represent three points that delineate the circumference of Earth's shadow, from where its radius can be calculated. 
Any good estimation of that radius allow us to move all the points up or down some $\epsilon$ distance to obtain a similar radius. If the initial set is moved to points $\{S_1',S_2',S_3'\}$,a circumference that passed for those points have a similar radius with an increment of  order of  $\epsilon$. On the other hand,  if the initial set is moved to points $\{S_1'',S_2'',S_3''\}$, a circunference can be constructed to pass for those points, with a radius with a decrement of  order of  $\epsilon$. The key part is that all constructed circunferences must be performed by using
points that were moved in the same direction, i.e., using the points from the same set. This maintains the estimation of $k_e$ within an error of $\epsilon$ order. The minimal errors are achieved when the chosen points of Earth's shadow are more or less in the middle of the Moon's circle. For images at the beginning or at the end of the duration of the lunar eclipse, the error of the estimation of the radius increases. 

\begin{figure}[h!]
\centering
\includegraphics[width=8cm]{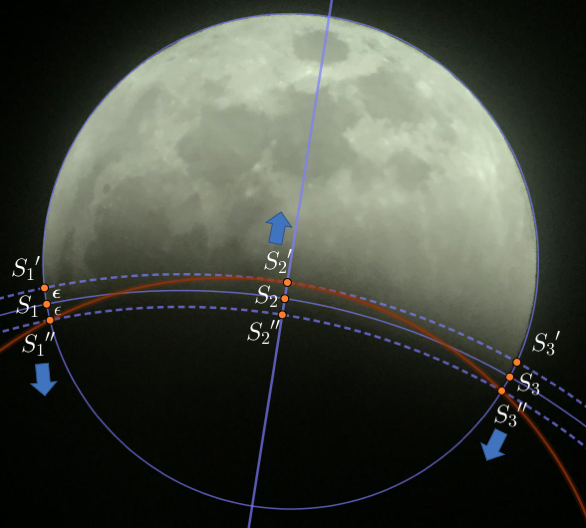}    
\caption{Example of bad choices for construction of circunferences using
points delineating the edge of Earth's shadow. The red circunference is constructed using three  points from different sets. This produces a large error in the calculation of radius of Earth's shadow. }
\label{limitation2}
\end{figure}

If this is not the case, then  a large error in radius estimation can be induced. This is shown in Fig.~\ref{limitation2}, where it is shown how a circumference constructed by using points from different sets leads to a very different radius (with an error much larger than $\epsilon$). In the figure, a circumference using points $\{S_1'',S_2',S_3''\}$ produces a very small circle. On contrary, if points $\{S_1',S_2'',S_3'\}$ were used, a very large circunference can be obtained. Both cases gives very bad estimation of the radius of Earth's shadow, and therefore, of $k_e$.

Finally, with all the these discussed considerations, we suggest that the chosen points must be  very separated from each other, with one of them in the middle. This implies that the best photographs are not when the lunar eclipse is in its totality, but when it is partial.

 \section{Conclusions}

The presented method is simple, straightforward and elegant enough to be presented as an activity proposal  in a physics or astronomy course when an eclipse occurs.
The results for $k$ are very impresive and close to the standard one given by the IAU. Furthermore, several other geometry problems can be derived from our photographs that a teacher can use in a more advanced course. 

It is convenient for students to take their own photographs to take advantage of their enthusiasm of the usage of smartphones. On the contrary, a group can take them with the help of the teacher and then explain the process to the rest of their classmates.

In order to  select the best photographs, the teacher could analyze some of the photos taken by the students according to the criteria presented along  this work. Then,  give to students some guidelines in order to reproduce more of them in the best way possible.
 On the other hand, in the case that taking photographs were not possible, the teacher can obtain images of lunar eclipses from internet, as there are several amateur astronomers who post photos of eclipses online. 
If the teacher decides to use these photographs, she/he must be careful to check if or how the photographs could have been manipulated in their dimensionality. Also, it must  checked if  the shadow have been treated in some way to increase its contrast, as this could significantly change the expected ratio.

Also, when this experience will be repeated for other eclipses, the teacher must consider that Earth-Moon and Sun-Earth distances will change, and that produces a slightly different value of $\mu$ of  Eq.~\eqref{valuofmu}. We refer to the software Stellarium \cite{estela} to obtain precise distances at the time of the eclipse.

The calculations presented here have very interesting mathematical elements that can be used depending on the level of the students. These  elements are not beyond of the reach of any elementary school student. Besides, the mathematical and geometrical tools used
make of this method a geometry and astronomy teaching tool for any student equipped with smartphone technology. In this way,  students are capable of taking this opportunity to
re--discover by themselves astronomical facts that appear in textboooks, and thus, to gain that knowledge in an active manner.

\begin{acknowledgments}
The authors express their gratitude to Francisco Rojas and to the referees for their valuable comments to our work.
H. C. thanks to  Grupo de Observaci\'on Astron\'omica of Universidad Adolfo Ib\'a\~nez. 
\end{acknowledgments}


\end{document}